\documentclass[11pt]{article}
\def\be{\begin{equation}}
\def\ee{\end{equation}}
\def\ba{\begin{eqnarray}}
\def\ea{\end{eqnarray}}
\def\la{\langle}
\def\ra{\rangle}
\def\a{\alpha}

\def\h{\hskip 1cm}

\def\lo{\longrightarrow}
\def\A1{A_{-1}}
\usepackage{epsfig}
\begin{document}
\begin{titlepage}
\vspace{4cm}
\begin{center}{\Large \bf A new family of matrix product states with Dzyaloshinski-Moriya interactions}\\
\vspace{2cm}\h Marzieh Asoudeh\\ \vspace{1cm} Department of
Physics, Shahid Beheshti
University, GC\\
19839-63113, Tehran, Iran\\
\end{center}

\vskip 2cm

\begin{abstract}
We define a new family of matrix product states which are exact
ground states of spin 1/2 Hamiltonians on one dimensional
lattices. This class of Hamiltonians contain both Heisenberg and
Dzyaloshinskii-Moriya interactions but at specified and not
arbitrary couplings.  We also compute in closed forms the one and
two-point functions and the explicit form of the ground state.
The degeneracy structure of the ground state is also discussed.

\end{abstract}
\vskip 2cm PACS Numbers: 03.67.HK, 05.40.Ca \hspace{.3in}
\end{titlepage}

\section{Introduction}\label{intro}
In the past few years, a lot of interest has been attracted to
the subject of matrix product states
\cite{mps1,mps3,mps3,mps4,mps4,mps5, mps6, mps7, mps8, mps9}.
There has been a revival of interest in this subject
\cite{qmps1,qmps2,qmps3,qmpsphase,qmps4,qmps5,qmps6} due to the
advances in quantum information theory \cite{niel} and the
techniques developed in this field. Quite recently the method of
matrix product states has been used for obtaining new models of
many body systems for which an exact ground state can be defined
and its many-body properties like the entanglement between various
sites can be calculated in closed form
\cite{aks1,aks2,aks3,akm1,akm2,abk,km,bt,a1}. This is interesting
in view of the fact that in some works, transitions in
entanglement behaviour has turned out \cite{oster,osb} to be a
good signal of
quantum phase transitions \cite{sach}. \\

The reason for this high level of interest is the complementary
role that the fields of condensed matter physics and quantum
information play in investigation of many body systems. The
matrix product formalism \cite{mpsoriginal1, aklt} is one of the
subjects which lies at the borderline of these two subjects. As
is well known, in this formalism, one starts from proposed states
whose expansion coefficients are the trace of product of given
matrices. While for numerical investigations, i.e. the density
matrix renormalization group (DMRG), one usually starts from
large dimensional matrices, to simulate ground states of given
Hamiltonians, in the approach which is used for finding exactly
solvable models, one starts from low dimensional matrices and
finds family of Hamiltonians for which these states are exact
ground states. This is the approach which has been used in many
of the works in the past few years \cite{
mps1,mps2,mps3,mps4,mps5,aks1,aks2,aks3,akm1,akm2,km,abk,bt,a1}. \\

In a recent work \cite{a1}, we classified all spin 1/2 matrix
product states which arise from two-dimensional matrices and
showed that some of them can undergo phase transitions. Doing the
same kind of classification for higher dimensional matrices turns
out to be a formidable task. A general classification in effect
requires classification of solutions of a system of non-linear
equations, which obviously is extremely difficult. However it is
possible to define new models by starting with special classes of
matrices which have nice algebraic properties. Such properties
allow us to bypass the above-mentioned difficult problem. The
price that we will pay is that we will obtain only restricted
family of models and not all models. However if we start from a
nice algebra, it is usually possible to end at physically
interesting models and by the subsequent analysis of these models
beyond the matrix product formalism, many useful information can
be obtained for such models which are otherwise difficult to
obtain. We will later see explicit examples of this method. \\

Following this method, in this paper we will introduce one such
family of matrix product states which is based on a simple algebra
\begin{equation}\label{algebra}
  XZ=\omega ZX
\end{equation}
where $\omega$ is a root of unity. We use a finite dimensional
representation of this algebra in which $Z$ and $X$ are
represented by generalized Pauli operators. Everything in the
model, from the state and its parent Hamiltonian (i.e. the
Hamiltonian which has the matrix product state as its ground
state) to the one- and two-point functions depend on the dimension
of the representation. We will show that while for $D=2$
dimensional representation of the algebra, the parent
Hamiltonian, is a one parameter of Heisenberg spin chain, for
$D\ne 2$ dimensional representation the Hamiltonian contains both
Heisenberg and Dzyaloshinskii-Moriya interactions \cite{D,M}, but
for specified values of couplings which depend on the dimensional
representation $D$. \\

It is to be noted that the matrix product state is by construction
a translation-invariant state and in some cases, it happens that
this state is a sum of two ground states each of which breaks the
translation symmetry. This is the case with Majumdar-Ghosh
\cite{mg1,mg2} model \cite{aks3}. Here we will see a similar
phenomena, where the matrix product state is the sum of many
ground states each of which breaks the translation symmetry of
the original
Hamiltonian.\\

The calculation of the one- and two-point functions is an
important step in this analysis. While such a calculation is
straightforward in the formalism of matrix product formalism, a
close scrutiny of their structure will give clues as to the whole
or part of the ground space structure and by following these
clues we can obtain valuable information about the degeneracy
structure of the ground state.\\

The structure of this paper is as follows. In section
(\ref{brief}) we give a brief introduction to matrix product
states, in section (\ref{model}) we introduce the representation
of the algebra (\ref{algebra}) and construct the family of matrix
product states and derive the parent Hamiltonian, in section
(\ref{corr}) we calculate the one and two point functions.
Finally we end up with a discussion.

\section{A brief introduction to matrix product
states}\label{brief}

First let us review the basics of matrix product states. Consider a
homogeneous ring of $N$ sites, where each site describes a $d-$level
state. The Hilbert space of each site is spanned by the basis
vectors $|i\ra, \ \ i=0,\cdots d-1$. A state
\begin{equation}\label{state}
    |\Psi\ra=\sum_{i_1,i_2,\cdots i_N}\psi_{i_1i_2\cdots
    i_N}|i_1,i_2,\cdots, i_N\ra
\end{equation}
is called a matrix product state if there exists $D$ dimensional
complex matrices  $A_i\in {C}^{D\times D},\ \ i=0\cdots d-1$ such
that
\begin{equation}\label{mat}
    \psi_{i_1,i_2,\cdots
    i_N}=\frac{1}{\sqrt{Z}}tr(A_{i_1}A_{i_2}\cdots A_{i_N}),
\end{equation}
where $Z$ is a normalization constant given by
\begin{equation}\label{z}
    z=tr(E^N)
\end{equation}
and
\begin{equation}\label{E}
E:=\sum_{i=0}^{d-1} A_i^*\otimes A_i.
\end{equation}
Here we are restricting ourselves to translationally invariant
states, by taking the matrices to be site-independent.

Let $O$ be any local operator acting on a single site. Then we can
obtain the one-point function on site $k$ of the chain $\la
\Psi|O(k)|\Psi\ra $ as follows:
\begin{equation}\label{1point}
    \la \Psi|O(k)|\Psi\ra = \frac{tr(E^{k-1}E_O E^{N-k})}{tr(E^N)},
\end{equation}
where
\begin{equation}\label{mpsop}
E_O:=\sum_{i,j=0}^{d-1}\la i|O|j\ra A_i^*\otimes A_j.
\end{equation}

The n-point functions can be obtained in a similar way. For example,
the two-point function $\la \Psi|O(k)O(l)|\Psi\ra$ can be obtained
as
\begin{equation}\label{2point}
\la \Psi|O(k)O(l)|\Psi\ra = \frac{tr(E_O(k)E_O(l)E^N)}{tr(E^N)}
\end{equation}
where $E_O(k):=E^{k-1}E_OE^{-k}$. Note that this is a formal
notation which allows us to write the n-point functions in a
uniform way, it does not require that $E$ is an invertible
matrix. Also by considering the permutation operator $P$ defined
as
\begin{equation}\label{2point}
P\mid\alpha\beta\ra=\mid\beta\alpha\ra
\end{equation}
and using equation(\ref{E})we see that $E^*=PEP$ and since $P^2=P$
we find that $Z^*=Z$, meaning that the normalization is real as
it should be. Also for a Hermitian operator $O$ in equation
(\ref{1point}) the one point function $ \la \Psi|O(k)|\Psi\ra$
will be real. The same reasoning applies for n-point functions.

\subsection{The Parent Hamiltonian}
Given a matrix product state, the reduced density matrix of $k$
consecutive sites is given by
\begin{equation}\label{rhok}
    \rho_{i_1\cdots i_k, j_1\cdots j_k}=\frac{tr((A_{i_1}^*\cdots A_{i_k}^*\otimes A_{j_1}\cdots A_{j_k})E^{N-k})}{tr(E^N)}.
\end{equation}
The null space of this reduced density matrix includes the solutions
of the following system of equations
\begin{equation}\label{cc}
    \sum_{j_1,\cdots, j_k=0}^{d-1}c_{j_1\cdots
    j_k}A_{j_1}\cdots A_{j_k}=0.
\end{equation}
Given that the matrices $A_i$ are of size $D\times D$, there are
$D^2$ equations with $d^k$ unknowns. Since there can be at most
$D^2$ independent equations, there are at least $d^k-D^2$ solutions
for this system of equations. Thus for the density matrix of $k$
sites to have a null space it is sufficient that the following
inequality holds
\begin{equation}\label{dD}
    d^k\ >\ D^2.
\end{equation}
Let the null space of the reduced density matrix be spanned by the
orthogonal vectors $|e_{\a}\ra, \ \ \ (\a=1, \cdots  s,\geq
d^k-D^2)$. Then we can construct the local hamiltonian acting on $k$
consecutive sites as
\begin{equation}\label{h}
    h:=\sum_{\a=1}^s \mu_{\a} |e_{\a}\ra\la e_{\a}|,
\end{equation}
where $\mu_{\a}$'s are positive constants. These parameters together
with the parameters of the vectors $|e_i\ra $ inherited from those
of the original matrices $A_i$, determine the total number of
coupling constants of the Hamiltonian.  If we call the embedding of
this local Hamiltonian into the sites $l$ to $l+k$ by $h_{l,l+k}$
then the full Hamiltonian on the chain is written as
\begin{equation}\label{H}
    H=\sum_{l=1}^N h_{l,l+k}.
\end{equation}
The state $|\Psi\ra$ is then a ground state of this hamiltonian with
vanishing energy. The reason is as follows:
\begin{equation}\label{Hrho}
\la \Psi|H|\Psi\ra=tr(H|\Psi\ra\la\Psi|)=\sum_{l=1}^N
tr(h_{l,l+k}\rho_{l,l+k})=0,
\end{equation}
where $\rho_{l,k+l}$ is the reduced density matrix of sites $l$ to
$l+k$ and in the last line we have used the fact that $h$ is
constructed from the null eigenvectors of $\rho$ for $k$ consecutive
sites. Given that $H$ is a positive operator, this proves the
assertion.\\

In view of the above introduction, we have a clear receipe for
constructing matrix product states and a family of parent
Hamiltonians. First one chooses the range of interaction $k$ and
then choose appropriate matrices, throwing away all spurious
degrees of freedom by appropriate transformations $A_i\lo S A_i
S^{-1} $ and reducing further the degrees of freedom by imposing
symmetries. In this way one ends with a reasonable set of matrix
product states, which hopefully may have applications in
description of real physical systems. In the following section,
we want to study a particular family of such states arising from
the simple algebra \ref{algebra}.

\section{The model}\label{model}

In this paper we specify the auxiliary matrices $A_0$ and $A_1$
as $D$ dimensional generalization of spin (Pauli) operators in
$x$ and $z$ directions. This is a D dimensional representation of
the algebra \ref{algebra}.  Hereafter, for convenience, we denote
$A_0$ and $A_1$ respectively by $X$ and $Z$. In the so called
computational basis for qudits, spanned by the orthonormal
vectors $\{|0\ra, |1\ra, \cdots |n\ra\}$ and we have
\begin{equation}\label{Xrep}
    X=\sum_{n=0}^{d-1} \mid n+1\ra \la n\mid,
\end{equation}
and
\begin{equation}\label{Zrep}
    Z=\sum_{n=0}^{D-1} \omega^n \mid n\ra \la n\mid,
\end{equation}
where $\omega=e^{\frac{2\pi i}{D}}$.\\
One finds from (\ref{Xrep} and \ref{Zrep}) that
\begin{equation}\label{xz}
ZX=\omega XZ.
\end{equation}

As explained in section{\ref{brief}}, equation (\ref{E}), the
transfer matrix for our model has the following simple form

\begin{equation}\label{transE}
E=X\otimes X+\overline{Z}\otimes Z.
\end{equation}

The matrix $E$ plays a central role for calculating the
normalization of the ground state and also in determining all the
correlation functions. We need to find the spectrum of this
matrix. The right and left eigenvectors of $E$ turn out to be
\begin{equation}\label{spect1}
   \mid\psi_k(r)\ra:=\frac{1}{\sqrt{D}}\sum_{n=0}^{D-1}\omega^{-rn}\mid
   n,n+k\ra,
\end{equation}

 \begin{equation}\label{spect2}
   \la\psi_k(r)\mid:=\frac{1}{\sqrt{D}}\sum_{n=0}^{D-1}\omega^{rn}\la
   n,n+k\mid,
\end{equation}
both of which correspond to the eigenvalue

\begin{equation}\label{eig}
\lambda_k(r)=\omega^k+\omega^r, \h k,r = 0, 1, \cdots D-1.
\end{equation}

It is easily checked that $\la \psi_k(r)|\psi_{k'}(r')\ra =
\delta_{k,k'}\delta_{r,r'}$.\\

Also we need to know $Z=tr(E^N)$, the normalization constant which
from (\ref{eig}) turns out to be
\begin{equation}\label{2}
 Z_D(N)= \sum_{k,r=0}^{D-1}(\omega^r+\omega^k)^N,
\end{equation}
where we have re-labeled the partition function as $Z_D(N)$ to
emphasize its dependence on the number of sites $N$ and the
dimension of the representation $D$.  From the binomial expansion,
we find

\begin{equation}\label{3}
   Z_D(N)=\sum_{r,k=0}^{D-1}\sum_{l=0}^{N}\left(\begin{array}{c}
                   N \\
                   l
                 \end{array} \right)\omega^{rl}\omega^{k(N-l)}
\end{equation}

Summing over $r$ and $k$ and noting that
$\sum_{r=0}^{D-1}\omega^{rl}=D\delta_{l,mD}$ and
$\sum_{k=0}^{D-1}\omega^{k(N-l)}=D\delta_{N-l,m'D}$ where $m$ and
$m'$ are integers we find that $Z_D(N)$ is nonzero only when $N$
is a multiple of $D$. Hereafter we assume that this is the case,
that is we set $N\lo ND$ in all the equations. The physical
necessity of this requirement will become clear when we discuss
the explicit form of the ground states.  With this modification,
we find the
\begin{equation}\label{zdelta}
   Z_D(DN)=D^2\sum_{l=0}^N(\begin{array}{c}
                   DN \\
                   Dl
                 \end{array}).
\end{equation}

This is a sum of $N$ terms, where $N$ is the number of sites and
as such is not a close expression. However we can use an identity
and rewrite it as a sum of $D$ terms which is much simpler. To
this end we consider the following binomial identity
\begin{equation}\label{simp1}
   (1+\omega^s)^{ND}=\sum_{l=0}^{ND}(\begin{array}{c}
                   ND \\
                   l
                 \end{array})\omega^{sl},
\end{equation}
and sum over both sides for $s=0, 1, \cdots D-1$ to obtain

\begin{eqnarray}\label{simp2}
   \sum_{s=0}^{D-1}(1+\omega^s)^{ND}&=&\sum_{s=0}^{D-1}\sum_{l=0}^{ND}(\begin{array}{c}
                   ND \\
                   l
                 \end{array})\omega^{sl}\cr
&=&\sum_{l=0}^{ND}(\begin{array}{c}
                   ND \\
                   l
                 \end{array})\sum_{s=0}^{D-1}\omega^{sl}
\end{eqnarray}
However, since $\omega^D=1$, the sum over $s$ in the right hand
side is non-vanishing only when $l$ is a multiple of $D$ and in
that case it will be equal to $D$. Therefore we find that

\begin{equation}\label{simp3}
   Z_{ND}=D\sum_{i=0}^{D-1
   }(1+\omega^s)^{ND}
\end{equation}
In contrast to (\ref{zdelta}), this is a sum over $D$ terms which
is
independent of the system size and obviously is much simpler.\\

For example for $D=2$ and for $D=3$ we have respectively

\begin{equation}\label{z2n}
   Z_{2N}=2^{2N+1}
\end{equation}

and

\begin{equation}\label{z3n}
   Z_{3N}=3(2^{3N}+2(-1)^N)
\end{equation}

We will now consider the parent Hamiltonian, for which the matrix
product state is an exact ground state.

\subsection{The Parent Hamiltonian}\label{hamiltonian}

To find the parent Hamiltonian, we use the relation $\sum
C_{ij}A_iA_j=0$ which in view of (\ref{algebra}) takes the form

\begin{equation}\label{null}
  C_{00}X^2+C_{01}XZ+C_{10}ZX+C_{11}Z^2=0.
\end{equation}

For $D\ne 2$ the operators $X^2$ and $Z^2$ are independent.
However in any dimension $ZX=\omega XZ$ and hence the vanishing
of the left hand side gives the following relations on the
coefficients
\begin{equation}\label{cc}
  C_{00}=C_{11}=0, \h C_{01}+\omega C_{10}=0.
\end{equation}
According to (\ref{cc}), these equations define the null space of
the density matrix of two adjacent sites. This null space is
spanned by only one vector, which we denote by $|e\ra$,
\begin{equation}\label{nulle}
|e\ra = |10\ra - \omega |01\ra.
\end{equation}
Thus the local Hamiltonian is given by

\begin{equation}\label{localh}
  h_{k,k+1}=|e\ra\la e|_{k,k+1} = (|10\ra-\omega|01\ra)(\la 10|-\overline{\omega}\la
  01|)_{k,k+1}.
\end{equation}
Writing the above expression in terms of the Pauli operators, we
find that
\begin{equation}\label{localhpauli}
  h_{k,k+1} = \frac{1}{2}(I-\sigma^z_{k}\sigma^{z}_{k+1}) -\omega
  \sigma^+_{k}\sigma^-_{k+1}-\overline{\omega}
  \sigma^-_{k}\sigma^+_{k+1}.
\end{equation}
After a simple re-scaling $H\lo 2H$ and neglecting an additive
constant, the full Hamiltonian becomes
\begin{equation}\label{HDne2}
H = \sum_{k=1}^N  \cos \frac{2\pi}{D}(\sigma^x_k\sigma^x_{k+1}+
\sigma^y_k\sigma^y_{k+1}) - \sigma^z_k \sigma^z_{k+1} + \sin
\frac{2\pi}{D}(\sigma^x_k\sigma^y_{k+1}-
\sigma^y_k\sigma^x_{k+1}).
\end{equation}

This is the general form of the Hamiltonian $H$ when we use the
$D\ne 2$ dimensional representation of the algebra
(\ref{algebra}). Therefore this Hamiltonian contains both
Heisenberg and Dzyaloshinskii-Moriya interaction \cite{D,M}, but
with specific and
not arbitrary couplings.\\

 In case we
use the $D=2$ dimensional representation, a different Hamiltonian
will be obtained, since in this case we have $X^2=Z^2=I$ and the
null-space condition (\ref{null}) will have a different solution.
In this null-space condition (\ref{null}) turns out to be
\begin{equation}\label{null}
  (C_{00}+C_{11})I+(C_{01}-C_{10})ZX=0,
\end{equation}
and the null space is spanned by two vectors, namely
\begin{equation}\label{e1e2}
  |e_1\ra = |00\ra - |11\ra, \h |e_2\ra = |01\ra+|10\ra.
\end{equation}
This larger null space means that the Hamiltonian has now two
arbitrary coupling constants, and by a re-scaling $H\lo 2H$, the
Hamiltonian takes the form
\begin{equation}\label{HD2}
   H=\sum_{i=1}^N(J_1-J_2)\left(\sigma^z_{k}\sigma^{z}_{k+1}-\sigma^x_{k}\sigma^x_{k+1}\right)+(J_1+J_2)(I+\sigma^y_{k}\sigma^y_{k+1}).
\end{equation}
In this case the Dzyaloshinskii-Moriya interaction is absent. Note
the Hamiltonian \ref{HD2} cannot be obtained from \ref{HDne2} in
a certain limit. Up to now the parent Hamiltonian has been
obtained for both types of representations of the matrix product
algebra (\ref{null}). The next step will be to calculate the
correlation functions. We limit ourselves to one and two-point
functions.

\section{One and two-point correlation functions}\label{corr}
The matrix product formalism allows a straightforward calculation
of correlation functions. According to (\ref{mpsop}), the matrix
product operators corresponding to the spin observables $\sigma_x,
\sigma_y$ and $\sigma_z$ are

\begin{eqnarray}\label{mpo}
  &&E_x := X\otimes Z + Z\otimes X, \cr
  &&E_y := -iX\otimes Z +i Z\otimes X, \cr &&E_z := X\otimes
X - Z\otimes Z.
\end{eqnarray}

For finding the average of Pauli operators in each site of the
chain we use equation (\ref{1point}). As explained in section
(\ref{brief}), for finite systems, this requires a
diagonalization of the transfer matrix $E$, while in the
thermodynamic limit even this problem simplifies to the
determination of the largest eigenvalue of the transfer matrix.
In some cases, i.e. for the two-dimensional representation of the
algebra (\ref{algebra}), one can even proceed in an alternative
and very simple way. Therefore we organize the content of this
section in two separate subsections, where we deal separately,
using different methods,  with 2-dimensional and higher
dimensional representations of the matrix product algebra.

\subsection{Two-dimensional representation}
Let us start by noting that for the two-dimensional
representation, we have $X^2=Z^2=I$ and $ZX=-XZ$. The same
relations hold with $X\lo Y$. These relations and the expression
for the transfer operator E given in (\ref{transE}) leads to the
simple relations
\begin{equation}\label{exey}
E_xE=E_yE=E_zE=0,
\end{equation}
from which we immediately obtain that
\begin{equation}\label{xy}
  \la \sigma^x_k\ra=\la \sigma^y_k\ra=\la \sigma^z_k\ra=0.
\end{equation}

It is obvious from the relations (\ref{exey}) that any correlation
function containing $\sigma_x$ and $\sigma_y$ operators on
non-adjacent sites will also vanish. For adjacent sites we have
to calculate traces of operators like $E_x^2E^{N-2}$, $E_y^2
E^{N-2}$ and $E_xE_yE^{N-2}$. It is simply obtained from
(\ref{mpo}) that

\begin{equation}\label{ex2ey2}
  E_x^2 E = 0, \h E_y^2 E = -4E, \h E_z^2E=0,
\end{equation}
which leads to
\begin{equation}\label{2point}
  \la \sigma^x_k\sigma^x_{k+1}\ra = 0 \h
  \la \sigma^y_k\sigma^y_{k+1}\ra =-4
  \frac{tr E^{2N-2}}{trE^{2N}}=-1,\h  \la \sigma^z_k\sigma^z_{k+1}\ra =
  0,
\end{equation}
where we have used equation (\ref{mpo}). The last relation implies
that the ground state is one in which the the spins of adjacent
sites are aligned in the $y$ directions but in opposite senses.
There are two such states and the matrix product state, which by
construction should be translation invariant, is the sum of these
two states, namely
\begin{equation}\label{psi2}
  |\Psi\ra=\frac{1}{\sqrt{2}}\left(|+_{y},-_{y}\ra^{\otimes N}+|-_{y},+_{y}\ra^{\otimes
  N}\right),
\end{equation}
where $|\pm_{y}\ra:=\frac{1}{\sqrt{2}}\left(\begin{array}{c} 1 \\
\pm i
\end{array}\right)$
are the two eigenstates of the $\sigma^y$ operator. By using the
relations $$\sigma^{z}|\pm_{y}\ra= |\mp_{y}\ra, \h
\sigma^{x}|\pm_{y}\ra=\pm i |\mp_{y}\ra, \h
\sigma^{y}|\pm_{y}\ra= \pm |\pm_{y}\ra, $$ one can readily check
that the two states in the superposition of $|\psi\ra$ are
actually annihilated by the Hamiltonian, so they are the ground
states of the Hamiltonian. In fact a simple way to see this, is
to rotate all the spins in the odd-numbered sites by an angle
$\pi$ around the $z$-axis. Under such a rotation, the Hamiltonian
(\ref{HD2}) transforms to

\begin{equation}\label{H2trans}
   H'=\sum_{i=1}^N(J_1-J_2)\left(\sigma^z_{k}\sigma^{z}_{k+1}+\sigma^x_{k}\sigma^x_{k+1}\right)+(J_1+J_2)(I-\sigma^y_{k}\sigma^y_{k+1}).
\end{equation}
or equivalently by adding and subtracting terms of the form
$\sigma^{y}_k\sigma^y_{k+1}$, it transforms to
\begin{equation}\label{H2trans}
   H'=\sum_{i=1}^N(J_1-J_2)\left(\vec{\sigma}_{k}\cdot\vec{\sigma}_{k+1}\right)-2J_1\sigma^y_{k}\sigma^y_{k+1},
\end{equation}
where $\vec{\sigma}_{k}\cdot\vec{\sigma}_{k+1}=
\sigma^x_{k}\sigma^{x}_{k+1}+\sigma^y_{k}\sigma^y_{k+1}+\sigma^z_{k}\sigma^{z}_{k+1}$
and we have dropped a total additive constant. In case that
$0<J_1<J_2$, the first term is a ferromagnetic interaction which
tends to align all the spins in one single direction, the
direction is determined by the second $y-y$ interaction. Hence we
have a doubly degenerate ground state of the form
$|\Phi_+\ra:=|y+\ra^{\otimes 2N}$ and $|\Phi_-\ra:=|y-\ra^{\otimes
2N}$ for $H'$. Rotating back all the odd-numbered spins by $\pi$
around the $z$ axis gives the two ground states of $H$ as
expressed in (\ref{psi2}),
 where their sum has been obtained as a
matrix product state. Thus in this case the MPS formalism guides
us through the complete degenerate structure of the ground state.
However if $J_1>J_2>0$, then the first term in $H'$ is an
anti-ferromagnetic interaction with a highly degenerate complex
structure. In this case the MPS only gives a very small part of
the spectrum.\\

\subsection{Higher dimensional representations}
We now turn to higher dimensional representation where the
Hamiltonian contains a Dzyialoshinski-Moriya interaction. In this
case the commutation relations do not allow a simple calculation
of the correlation functions for finite systems. Therefore we
calculate the correlations directly in the thermodynamic limit.
In this limit, only the largest eigenvalue and eigenvector of the
transition operator $E$ will survive the limit $N\lo \infty$ and
one finds that

\begin{equation}\label{thermo}
  \la O_k\ra = \frac{1}{\lambda_{max}}\la
  \lambda_{max}|E_O|\lambda_{max}\ra,
\end{equation}
and

\begin{equation}\label{thermo}
  \la O_kO_{k+1}\ra = \frac{1}{\lambda_{max}^2}\la
  \lambda_{max}|E_O^2|\lambda_{max}\ra,
\end{equation}

where $\lambda_{max}$ and $|\lambda_{max}\ra$ are the largest
eignevalue of $E$ and its corresponding eigenvector. From
(\ref{eig}) we know that the largest eigenvalue of $E$ is $2$
corresponding to the generalized Bell state
$|\Psi_{00}\ra=\frac{1}{\sqrt{D}}\sum_{n=0}|n,n\ra$. For the
one-point functions we need the matrix elements of the operators
$E_x, E_y,$ and $ E_z$ on this Bell state. It is obvious from
(\ref{mpo}) that $\la \psi_{00}|E_x|\psi_{00}\ra =\la
\psi_{00}|E_x|\psi_{00}\ra =0 $. For $E_z$ we find that

\begin{eqnarray}\label{ez}
  \la \psi_{00}|E_z|\psi_{00}\ra &=& \frac{1}{D}\sum_{n,m=0}^{D-1}\la
  m,m|X\otimes X - \overline{Z}\otimes Z|n,n\ra \cr&=&
  \frac{1}{D}\sum_{n,m=0}^{D-1} (\delta_{m,n+1}-\delta_{m,n})=0.
\end{eqnarray}

Therefore we find
\begin{equation}\label{xyz}
  \la \sigma_x\ra=\la \sigma_y\ra = \la \sigma_z\ra=0.
\end{equation}

We now proceed to calculate the two-point functions for adjacent
sites. To do this we need the following matrix elements which are
readily obtained from (\ref{mpo}) after simple manipulations:

\begin{eqnarray}\label{matrixelements1}
\la \psi_{00}|E_x^2 |\psi_{00}\ra &=&  \omega + \omega^{-1} = 2
\cos \frac{2\pi}{D},\cr \la \psi_{00}|E_y^2 |\psi_{00}\ra &=&
\omega + \omega^{-1} = 2 \cos \frac{2\pi}{D},\cr \la
\psi_{00}|E_z^2 |\psi_{00}\ra &=& 0,
\end{eqnarray}
and
\begin{eqnarray}\label{matrixelements2}
\la \psi_{00}|E_xE_y |\psi_{00}\ra &=& i( \omega - \omega^{-1})
=-2 \sin \frac{2\pi}{D},\cr \la \psi_{00}|E_yE_x |\psi_{00}\ra
&=& -i(\omega - \omega^{-1}) = 2 \sin \frac{2\pi}{D}.
\end{eqnarray}

Moreover all the other quadratic matrix elements turn out to
vanish. From the above matrix elements, one can easily obtain the
correlation functions from (\ref{thermo}) as follows:

\begin{equation}\label{corr}
\la \sigma^x_k\sigma^x_{k+1}\ra = \la
\sigma^y_k\sigma^y_{k+1}\ra=\frac{1}{2}\cos \frac{2\pi}{D}\ \ \ \
,  \ \ \ \ \la \sigma^z_k\sigma^z_{k+1}\ra=0,
\end{equation}
and
\begin{equation}\label{corr}
\la \sigma^x_k\sigma^y_{k+1}\ra = -\la
\sigma^y_k\sigma^x_{k+1}\ra=\frac{1}{2}\sin \frac{2\pi}{D}.
\end{equation}
All the other two-point correlation functions vanish for nearest
neighbor sites. The above results can be written in the following
compact form:

\begin{equation}\label{compact}
  \la \vec{\sigma}_k\cdot \vec{\sigma}_{k+1}\ra =
  \cos\frac{2\pi}{D}\h \la \vec{\sigma}_k\times \vec{\sigma}_{k+1}\ra =
  -\sin\frac{2\pi}{D} \hat{z}.
\end{equation}

From these correlation functions along with the result
(\ref{psi2}) which we obtained for the explicit form of the
ground state for the two-dimensional representations, we can guess
an explicit form for the ground state of the Hamiltonian
(\ref{HDne2}). We will later check the validity of our guess by
direct calculations.  The ground state (\ref{psi2}), turned out to
be a superposition of two product states, each one of them was
nothing but a juxtaposition of spins in opposite directions. In
fact each of the two product states are ground states of the
Hamiltonian separately, and the matrix product state superposes
them to ensure its own translational invariance. The same thing
may happen here, that is we may find product states each of which
is a ground state of the Hamiltonian and the matrix product state
is a superposition of them. Inspired by the form of the ground
states for the D=2 representation, we construct a product state
which gives the same form of correlation functions as in
(\ref{compact}). This construction is a generalization of the
states in (\ref{psi2}). Let
$$|\phi\ra := \sin\phi|+_x\ra+  \cos\phi|+_y\ra,$$ where $|+_x\ra$ and $|+_y\ra$ are the spin states in the positive $x$ and $y$ directions respectively.
We now form the following product state on $D$ consecutive sites
(instead of two sites in the $D=2$ case)

\begin{equation}\label{phistate}
 |\chi(\phi)\ra:= |\phi\ra|\phi-\frac{2\pi}{D}\ra|\phi-\frac{4\pi}{D}\ra\cdots|\phi-\frac{2(D-1)\pi}{D}\ra.
\end{equation}
Each consecutive spin has been rotated anti-clockwise by an angle
of $\frac{2\pi}{D}$ around the $z$ axis, in analogy with the
$D=2$ dimensional case. That is, the spins in each block of $D$
consecutive sites are distributed regularly around a circle in
the $x-y$ plane. It is obvious that for any such block state,
comprising $D$ consecutive sites, the average of spin operator in
any direction should vanish, that is
\begin{figure}
  \centering
  \epsfig{file=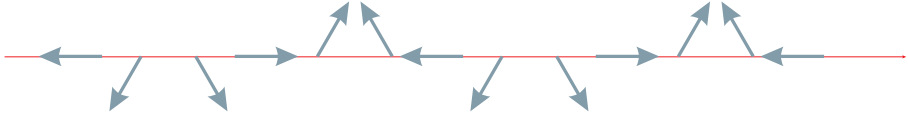,width=12cm}
  \caption{(Color online) One of the product states $|\chi\ra$ (for D=6). The matrix product state is a superposition of $D$ such states, each shifted by one site with respect to the previous one. }
  \label{min-ent-1}
\end{figure}

\begin{equation}\label{averagechi}
  \la \sigma^{a}\ra_{\chi(\phi)}:= \frac{1}{D}\sum_{i=0}^{D-1}\la
  \chi(\phi)|
  \sigma^{a}_i|\chi(\phi)\ra=0,\ \ \ \ \ a=x,y,z.
\end{equation}
The reason is nothing but the regular distribution of the spin
states $|\phi\ra$ over the circle. The same type of reasoning,
which can be supported by direct and explicit calculations, shows
the validity of the two-point correlations (\ref{compact}) for any
block state
of the above form.\\

We can extend such a state to the whole lattice as a
non-translation invariant $|\chi(\phi)\ra^{\otimes N}$, which
obviously has the same kinds of correlations as in
(\ref{compact}). The matrix product state, being translation
invariant is nothing but the following superposition
\begin{equation}\label{MPSD}
  |\Psi_D\ra:=|\chi(0)\ra^{\otimes N}+ |\chi(\frac{2\pi}{D})\ra^{\otimes N}+ |\chi(\frac{4\pi}{D})\ra^{\otimes N}\cdots
 + |\chi(\frac{2(D-1)\pi}{D})\ra^{\otimes N}.
\end{equation}
This formula for the MPS ground state generalized the explicit
formula (\ref{psi2}) to higher dimensional representations of the
algebra.\\

It is instructive to know the origin of the character of the
ground state (\ref{MPSD}). Let us proceed as in the $D=2$ case and
rotate the spins individually, but this time, according to a
different pattern. Consider the local Hamiltonian

\be\label{localH} h_{k,k+1} =  \cos
\frac{2\pi}{D}(\sigma^x_k\sigma^x_{k+1}+
\sigma^y_k\sigma^y_{k+1}) - \sigma^z_k \sigma^z_{k+1} + \sin
\frac{2\pi}{D}(\sigma^x_k\sigma^y_{k+1}-
\sigma^y_k\sigma^x_{k+1})\ee and rotate the spin at site $k+1$ by
an angle $\theta=-\frac{2\pi}{D}$ around the $z$ axis. Under this
transformation, the spin operators transform as follows:

\begin{eqnarray}\label{Drot}
\sigma^z_{k+1}&\lo& \sigma^z_{k+1}\cr \sigma^x_{k+1}&\lo& -\cos
\theta \ \sigma^x_{k+1}+\sin \theta\ \sigma^y_{k+1}\cr
\sigma^y_{k+1}&\lo& \ \ \ \sin \theta \ \sigma^x_{k+1}-\cos
\theta\ \sigma^y_{k+1}.
\end{eqnarray}

Under this transformation, it is readily seen that $h_{k,k+1}$
transforms to an isotropic Heisenberg ferromagnet,
$h^{Heis}_{k,k+1}$

\begin{equation}\label{hHeis}
  h^{Heis}_{k,k+1}=-
  \left(\sigma^x_k\sigma^x_{k+1}+\sigma^y_k\sigma^y_{k+1}+\sigma^z_k\sigma^z_{k+1}\right).
\end{equation}
It is well known that the ground state of the Heisenberg
Hamiltonian $H^{Heis}=\sum_{k}h^{Heis}_{k,k+1}$ is the state in
which all the spins align in one direction, and since the
Hamiltonian has rotational symmetry, the ground state consists of
a whole spin multiplet, which is the multiplet with the Highest
spin. For a lattice with $ND$ sites, this multiplet has the spin
$\frac{ND}{2}$ whose top state is
$|\frac{ND}{2},\frac{ND}{2}\ra=|z_+\ra^{\otimes ND}$. The other
states are obtained by the action of the angular momentum operator
$L_-:=\sum_{k=1}^{ND}\sigma^-_k$ on this top state. That is, the
other un-normalized ground states are
\begin{equation}\label{other}
  |\frac{ND}{2},m\ra = L_-^{\otimes
  (\frac{ND}{2}-m)}|\frac{ND}{2},\frac{ND}{2}\ra.
\end{equation}
Acting on these states by the inverse rotations $\bigotimes_{k}
R_{z}(\frac{2k\pi }{D})$ we obtain the ground states of the
original Hamiltonian in the form

\begin{equation}\label{other}
  \Psi_{m}:=\left[\bigotimes_{k}
R_{z}(\frac{2k\pi }{D})\right] L_-^{\otimes
  (\frac{ND}{2}-m)}|\frac{ND}{2},\frac{ND}{2}\ra.
\end{equation}

The ground states in $|\chi(\phi)\ra^{\otimes N}$ from which the
matrix product state (\ref{MPSD}) is constructed are only part of
the ground space of the Hamiltonian. As expected in the matrix
product formalism one starts from one single state, but a careful
analysis of that single state and its correlation functions,
assisted by some guesswork and physical analogies, can unravel
the whole degeneracy structure of the Hamiltonian.

\section{Discussion}
We have started from a simple matrix product algebra, namely $X\
 Z=\omega\ Z \ X$ where $\omega^D=1$ and proceeded to obtain the
properties of the matrix product state which corresponds to this
algebra. There is a distinct difference between the cases $D=2$
and $D\ne 2$. In the former case, the model describes a system of
spins on a line with Heisenberg type interaction but with
prescribed couplings, in the latter case, the model describes
both a Heisenberg and a Dzyaloshinski-Morya (DM) interaction. In
both cases we have been able to calculate the one and two point
functions and from the insight that these functions have
provided, we have been able to obtain a more thorough
understanding of these models, including the degeneracy structure
of their ground state. One possible line of expansion of this
work is to start from a Heisenberg ferromagnetic chain of
arbitrary spins in a magnetic field. If now we rotate in a
periodic pattern the magnetic fields and demand, similar to the
work of Kurman et al \cite{kur}, that all the cross terms except
the Heisenberg and the DM interactions vanish, we will end up
with a new form of exactly solved Heisenberg model in which DM
interactions are also present. The difficulty with this approach
is that if we start with a uniform magnetic field, the final
magnetic fields will not be uniform anymore, due to the periodic
rotations. On the other hand if we start with a non-uniform
magnetic field and demand that the final magnetic field will be
uniform along the chain, then the spectrum of the original
Hamiltonian will be very difficult to find. We hope to overcome
these difficulties in our future work.\\

{\textbf{Acknowledgements}} I would like to thank V. Karimipour
for many instructive discussions and valuable comments and also
for a critical reading of the manuscript.

{}


\begin{thebibliography}{}


\bibitem{mps1}  A. Klumper, A. Schadschneider and J. Zittartz, J. Phys. A
(1991) L293; Z. Phys. B, 87 (1992) 281; Europhys. Lett., 24
(1993) 293.

\bibitem{mps2}  H. Niggemann, and J. Zittartz, J. Phys. A: Math. Gen. 31,
9819-9828 (1998).

\bibitem{mps3}  E. Bartel, A. Schadschneider and J. Zittartz, Eur. Phys.
Jour. B, 31, 2, 209-216 (2003). 14

\bibitem{mps4}  M. A. Ahrens, A. Schadschneider, and J. Zittartz, Europhys.
Lett. 59 6, 889 (2002).

\bibitem{mps5}  A. K. Kolezhuk and H. J. Mikeska, Phys. Rev. Lett. 80, 2709
(1998); Int. J. Mod. Phys. B, 12, 2325-2348 (1998).

\bibitem{mps6}  A. K. Kolezhuk, H. J. Mikeska, and Shoji Yamamoto, Phys. Rev.
B 55, R3336 - R3339 (1997).

\bibitem{mps7}  H. Niggemann, A. Kl¨umper, and J. Zittartz, Eur. Phys. J. B
13, 15 (2000).

\bibitem{mps8}  H. Niggemann, A. Kl¨umper, and J. Zittartz, Z. Phys. B 104,
103 (1997).

\bibitem{mps9}  A. Kl¨umper, S. Matveenko, and J. Zittartz, Z. Phys. B 96,
401 (1995).


\bibitem{qmps1}  F. Verstraete, J. J. Garcia-Ripoll, and J. I. Cirac, Phys.
Rev. Lett. 93, 207204 (2004).

\bibitem{qmps2}  F. Verstraete, J. I. Cirac, Phys. Rev. B 73, 094423 (2006);
T. J. Osborne, Phys. Rev. Lett. 97, 157202 (2006); M. B.
Hastings, Phys. Rev. B 73, 085115 (2006).

\bibitem{qmps3}  G. Vidal, Phys. Rev. Lett. 91, 147902 (2003).

\bibitem{qmpsphase}  M. M. Wolf, G. Ortiz, F. Verstraete and I. Cirac, Phys. Rev.
Lett. 97, 110403 (2006).

\bibitem{qmps4}  F. Verstraete, M. A. Martin-Delgado, and J. I. Cirac, Phys.
Rev. Lett. 92, 087201 (2004); F. Verstraete, M. Popp, and J. I.
Cirac, Phys. Rev. Lett. 92, 027901 (2004); F. Verstraete, J. I.
Cirac, J. I. Latorre, E. Rico, and M. M. Wolf, Phys. Rev. Lett. 94
140601(2005).


\bibitem{qmps5}  F. Verstraete, D. Porras, J. I. Cirac, Phys. Rev. Lett. 93,
227205 (2004).

\bibitem{qmps6}  J. M. Roman, G. Sierra, J. Dukelsky, and M. A.
Martn-Delgado, J. Phys. A: Mathematical and General, 31, 48,
9729-9759(1998).

\bibitem{niel} M. A. Nielsen, and I. L. Chuang;Quantum computation and
quantum information, Cambridge University Press, Cambridge, 2000.


\bibitem{aks1}  M. Asoudeh, V. Karimipour, and A. Sadrolashrafi, Phys. Rev.
A 76, 012320 (2007).

\bibitem{aks2}  M. Asoudeh, V. Karimipour and A. Sadrolashrafi, Phys. Rev.
B, 75, 224427 (2007).

\bibitem{aks3}  M. Asoudeh, V. Karimipour, and A. Sadrolashrafi, Phys. Rev.
A 76, 012320 (2007).

\bibitem{akm1}  S. Alipour, V. Karimipour and L. Memarzadeh, Phys. Rev. A
75, 052322 (2007).


\bibitem{akm2}  S. Alipour, V. Karimipour, and L. Memarzadeh, Eur. Phys. J.
B 62, 159-169 (2008).


\bibitem{abk}  S. Alipour, S. Baghbanzadeh, and V. Karimipour, C Europhysics
Letters (EPL), 84 (2008) 67006.

\bibitem{km}  V. Karimipour and L. Memarzadeh, Phys. Rev. B 77, 094416
(2008). 15


\bibitem{bt}  A. Trebedi and I. Bose, Phys. Rev. A 75, 042304(2007).


\bibitem{a1}  M. Asoudeh, Physica A; 389, 8, 1555-1564 (2010).

\bibitem{D} I. Dzyaloshinskii, J. Phys. Chem. Solids 4, 241 (1958).

\bibitem{M} T. Moriya, Phys. Rev 120, 91 (1960).


\bibitem{sach}  S. Sachdev, Quantum Phase Transitions (Cambridge University
Press, Cambridge, 1999).

\bibitem{oster}  A. Osterloh, L. Amico, G. Falci, and R. Fazio, Nature 416, 608 (2002).

\bibitem{osb}  T. J. Osborne and  M. A. Nielsen, Phys. Rev. A 66,
032110 (2002).


\bibitem{mpsoriginal1}  M. Fannes, B. Nachtergaele and R. F. Werner, Commun. Math.
Phys. 144, 443 (1992).

\bibitem{aklt}  I. Affleck, T. Kennedy, E. H. Lieb, H. Tasaki, Commun.Math.
Phys. 115, 477 (1988);


\bibitem{mg1} C. K. Majumdar and D. P. Ghosh, J. Math. Phys. 10
(1969)1388.

\bibitem{mg2} C. K. Majumdar and D. P. Ghosh, J. Math. Phys. 10
(1969)1399.


\bibitem{kur} J. Kurmann, H. Thomas and G. Muller, Physica 112 A
235(1982).



\end{thebibliography}
\end{document}